\documentclass[%
 reprint,
 amsmath,amssymb,
 aps,
]{revtex4-2}

\usepackage{graphicx}
\usepackage{dcolumn}
\usepackage{bm}
\usepackage{lipsum}
\usepackage{subfigure}
\begin{document}

\preprint{APS/123-QED}

\title{Solving the Spectral Problem via the Periodic Boundary Approximation in $\phi^6$ Theory
}


\author{Lingxiao Long}
\affiliation{School of Space Science and Physics, Shandong University at Weihai, 264209, China}

\author{Yunguo Jiang}
\email{jiangyg@sdu.edu.cn}
\affiliation{School of Space Science and Physics, Shandong University at Weihai, 264209, China}
\affiliation{Shandong Provincial Key Laboratory of Optical Astronomy and Solar-Terrestrial Environment, Institute of Space Sciences, Shandong University, Weihai, 264209, China}

\date{\today}

\begin{abstract}


In $\phi^6$ theory, the resonance scattering structure is triggered by the so-calls delocalized modes trapped between the $\bar{K}K$ pair. The frequencies and configurations of such modes depend on the $\bar{K}K$ half-separation 2$a$, can be derived from the Schr\"{o}dinger-like equation. We propose to use the periodic boundary conditions to connect the localized and delocalized modes, and use periodic boundary approximation (PBA) to solve the spectrum analytically. In detail, we derive the explicit form of frequencies, configurations and spectral wall locations of the delocalized modes. We test the analytical prediction with the numerical simulation of the Schr\"{o}dinger-like equation, and obtain astonishing agreement between them at the long separation regime. 

\end{abstract}

\maketitle


\section{\label{sec:level1}Introduction}
Solitons in nonintergrable systems have been studied for many decades, with its application in condensed matter physics \cite{CON,CON2}, field theory \cite{F1,F2,F3}, and cosmology \cite{COS1}. Kink, as one of the typical kinds of solitons in 1+1 dimensions, is completely defined by a global invariant called topological charge. An intricate case is kink-antikink collision, with its mechanism of the reversible particle-like energy transfer between translation and inner vibration. The essential condition of multi-bounce resonances in general kink-antikink scattering, is the existence of an internal shape(vibration) mode \cite{campbell,Goodman1,Manton1}, which represents internal excitation and their configurations can be derived from linear perturbation theory. One non-intergrable model is $\phi^4$ model in which a kink can possess a shape mode.  We call such mode localized mode belonging to a single kink. However, localized modes are invalid in higher order polynomial theories. In prototypical $\phi^6$ theory, although a single kink solution don't possess such one mode, delocalized modes exist considering the superposition of static antikink and kink in linear perturbation theory. These delocalized modes are responsible for the emergence of resonance structures. Furthermore, the quasinormal mode also has effect on resonance structure, which is similar to the shape mode but has dissipation tails leading to energy leakage \cite{Dorey:qnm}.

 Generally, the way to derive delocalized modes, quasinormal modes and their frequencies is spectral problem. The spectral problem in $\phi^6$ was first noted in \cite{dorey2011kink} and recently investigated in \cite{Adamrev6}. Moreover, the spectral wall phenomena in $\phi^6$ theory are further studied in \cite{AdamSpe6}. However, according to the complexity of the Schr\"{o}dinger potential in linear perturbation theory, their methods to derive the spectral structure are numerical. To find analytical forms of shape modes and their frequencies is an stepping stone to construct effective models in the future. 

In our letter, we propose an analytical solution of the excited spectrum in $\phi^6$ theory and find a bridge connected the localized and delocalized modes. Crucially, we test our analytical predictions with full numerical solutions of the Schr\"{o}dinger-like equation. In the asymptotic large separation regime  (nearly $a>2$, \emph{a} represents the half separation between the kink and antikink), the analytical prediction has astonishing agreements with the numerical results  in tests of either frequency or delocalized modes. 

\section{\label{sec:level2}Dissipation modes of a single kink}
The perturbation problem of the single kink is already solved by \cite{selfex}. When the antikink and kink separates, local Schr\"{o}dinger potential near the kink part approximately equals that of a single kink, and the antikink part satisfies the similar condition. The essence of our method, is using the periodic boundary condition to combine the solutions of kink and antikink parts, namely the periodic boundary approximation (PBA) method.
 In this section, we introduce the dissipation modes of the single kink, which play an important role in connecting the localized and delocalized modes. 

\begin{figure}[htp]
    \centering
    \includegraphics[width=7cm]{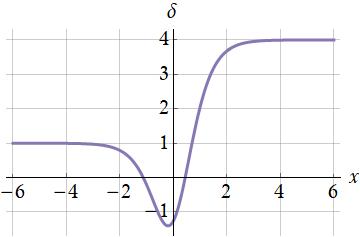}
    \caption{The plot of the Schr\"{o}dinger potential $U(x)$}
    \label{pot}
\end{figure}

In $\phi^6$ theory, the Lagrangian density in 1+1 dimensions is defined as \cite{Gani}
\begin{equation}\label{eqn0} 
\mathcal L=\frac{1}{2}\partial_\mu\phi\partial^\mu\phi-\frac{1}{2}\phi^2(\phi^2-1)^2,
\end{equation}
and the Euler-Lagrange equation is
\begin{equation}\label{eqn1.1} 
\frac{\partial^2 \phi}{\partial t^2}-\frac{\partial^2 \phi}{\partial x^2}+\phi(\phi^2-1)^2+2\phi^3(\phi^2-1)=0.
\end{equation}
It is widely known that the kink soliton, is a non-trivial solution of Eq. (\ref{eqn1.1}) interpolating between two adjacent vacua. The four $\phi^6$ kink solutions are
\begin{equation}\label{eqn1.2} 
\begin{split}
&\phi_{(0,1)}=\sqrt{(1+\tanh x)/2},\\
&\phi_{(1,0)}=\sqrt{(1-\tanh x)/2},\\
&\phi_{(0,-1)}=-\sqrt{(1+\tanh x)/2},\\
&\phi_{(-1,0)}=-\sqrt{(1-\tanh x)/2}.
\end{split}
\end{equation}
By setting $\phi=\phi_{(0,1)}+\eta(x)e^{i\omega t}$, we use perturbation methods to derive the  self-excited modes. The corresponding Schr\"{o}dinger-like equation can be expressed as \cite{dorey2011kink}
\begin{equation}\label{eqn2} 
\eta_{xx}=[U(x)-\omega^2] \eta,
\end{equation}
where $U(x)=15\phi_{(0,1)}^4-12\phi_{(0,1)}^2+1$ is the potential (See FIG.\ref{pot}). 
This kind of potential well is too narrow to contain more than one bound state \cite{selfex}. So except for the translation mode with $\omega=0$, it merely has  continuous spectrum when $1\leq\omega<2$, and the configurations of dissipation solutions can be expressed as \cite{selfex}
\begin{equation}\label{eqn1}
\begin{split}
\delta_{D}=N\frac{e^{-b^{*}x}}{(e^{-x}+e^{x})^b} F(b-\frac{3}{2},b+\frac{5}{2},k_{+}+1,\frac{e^{-x}}{e^{-x}+e^x}),\\
\end{split}
\end{equation}
where $k_{+}=\sqrt{4-\omega^2},\,\,k_{-}=\sqrt{\omega^2-1},\,\,b=(k_{+}-ik_{-})/2$,
\emph{N} is a normalization real constant and $F$ is the hypergeometric function.
$\delta_{D}(x)$ has the asymptotic behavior, which is given as  \cite{selfex}
\begin{equation}\label{eqn3}
\begin{split}
&\delta_{D}(x)\rightarrow 0\,,\,\,\,\qquad\qquad\qquad\qquad\qquad x\rightarrow+\infty,\\&
\delta_{D}(x)\rightarrow N(Ae^{ik_{-}x}+A^*e^{-ik_{-}x})\,,\,\,\,\,\,\, x\rightarrow-\infty,
\end{split}
\end{equation}
where 
\begin{equation}\label{eqn4}
\begin{split}
A=\frac{\Gamma(k_{+}+1)\Gamma(-ik_{-})}{\Gamma(b-\frac{3}{2})\Gamma(b+\frac{5}{2})},
\end{split}
\end{equation}
and $\Gamma$ is Gamma function.

The form of asymptotic behavior (\ref{eqn3}) means that $\delta_{D}(x)$ has specific characteristic.  On the one side it is bound, but on the other side it oscillates towards the infinity. 

\section{\label{sec:level3}Antikink-kink spectrum}
 
\begin{figure}[htbp]
    \centering
        \subfigure[]{
    	\includegraphics[width=9cm]{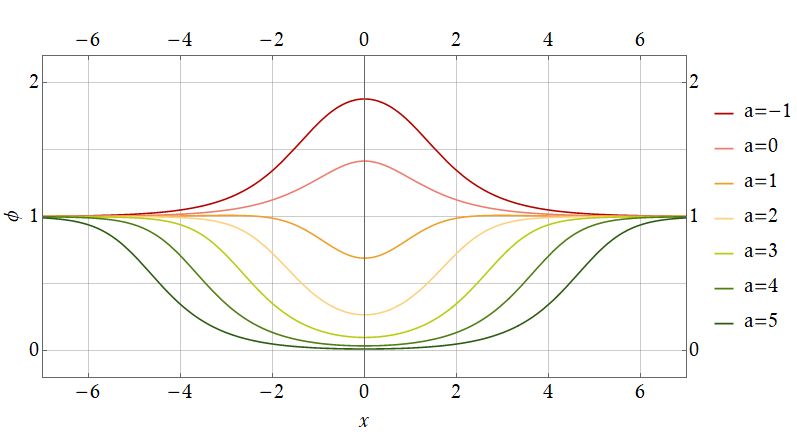}
        \label{AKK}
    }
    \quad    
    \subfigure[]{
        \includegraphics[width=9cm]{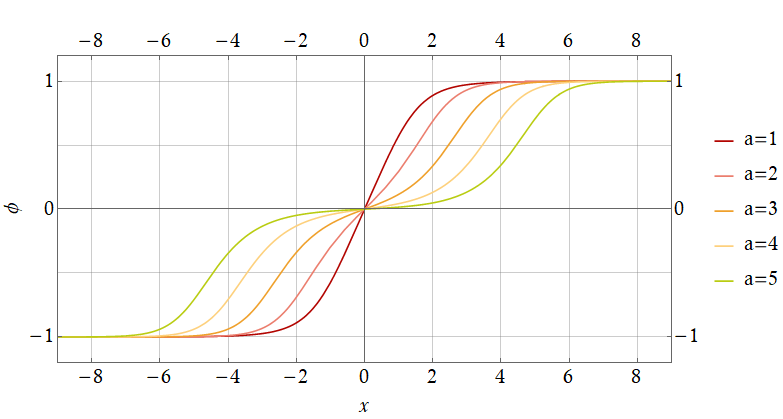}
        \label{KK}
    }
    \quad
    \caption{Examples of the naive superposition of $\bar{K}K$ (top) and $KK$ (bottom).}
    \label{K}
\end{figure}

In this section, we use PBA to derive the spectrum in the $\bar{K}K$ collision (See FIG.\ref{AKK}). In detail, we study the spectral structure, locations of spectral walls and configurations of shape (delocalized) modes. Due to the presence of the y-axis symmetry, we add a period boundary condition on the dissipation modes of the single kink. So that the configuration of the delocalized mode can be approximated by the combination of the selected dissipation mode and its mirror image for the y-axis. 
Since the configuration of the excitation is an even function, so the derivative at the centre of the figuration is zero.

\begin{figure}[htbp]
    \centering
        \subfigure[]{
    	\includegraphics[width=8.2cm]{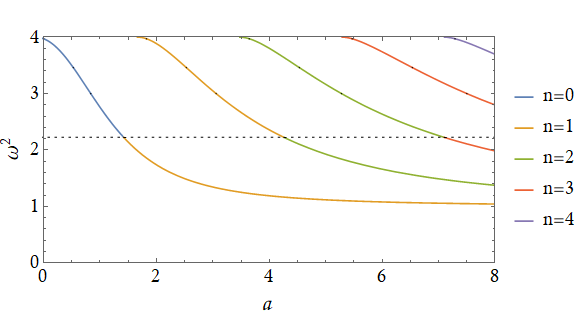}
        \label{21}
    }
    \quad    
    \subfigure[]{
        \includegraphics[width=8.7cm]{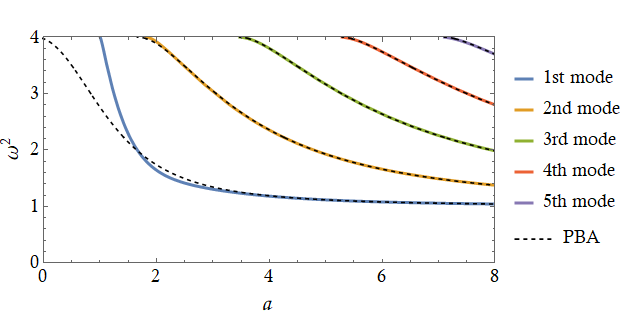}
        \label{22}
    }
    \quad
    \caption{In the $\bar{K}K$ collision: (a) Spectral structure via PBA, different colors represent curves with different \emph{n}; (b) the black dashed lines denote the merge of the colored curves in panel (a), The colored lines denote the result of the Schr\"{o}dinger-like equation.}
    \label{DR}
\end{figure}

Consider the  $\phi_{(0,1)}$ kink locating at $x=a$, the period boundary condition at $x=0$ can be expressed by
\begin{equation}\label{eqn4.1}
\begin{split}
&\frac{\partial\delta_{D}(x-a)}{\partial x}\bigg|_{x = 0} =0.
\end{split}
\end{equation}
For $\phi^6$ theory, when $\emph{a}>2$, the kink and antikink separate almost completely. This half separation is large enough that we can use the asymptotic approximation in Eq.(\ref{eqn3}). Now Eq. (\ref{eqn4.1}) takes the form as
\begin{equation}\label{eqn4.2}
\begin{split}
&\frac{\partial{{\rm Re}[A(\omega)e^{ik_{-}(x-a)}]}}{\partial x}\bigg|_{x = 0} =0.
\end{split}
\end{equation}
It equals to
\begin{equation}\label{eqn4.3}
\begin{split}
&\frac{\partial{[{\rm Re}(A(\omega))\cos(k_{-}(x-a))]}}{\partial x}\bigg|_{x = 0} =\\&
\frac{\partial{[{\rm Im}(A(\omega))\sin(k_{-}(x-a))]}}{\partial x}\bigg|_{x = 0},
\end{split}
\end{equation}
then
\begin{equation}\label{eqn4.3}
\begin{split}
-k_{-}(\omega)a+\arctan[\frac{{\rm Im}(A(\omega))}{{\rm Re}(A(\omega))}]=-n\pi.
\end{split}
\end{equation}
where \emph{n} is a non-negative integer. Finally we derive the  half separation \emph{a} as
\begin{equation}\label{eqn5}
\begin{split}
a=\frac{1}{k_{-}(\omega)}[\arctan(\frac{{\rm Im}(A(\omega))}{{\rm Re}(A(\omega))})+n\pi].
\end{split}
\end{equation}
Based on this equation, we plot the spectral structure ($\omega^2$ versus $a$) in FIG. \ref{21} for $n=0,1,2,3$ and $4$. In FIG. \ref{22}, we also plot the spectral structure from the numerical result of the Schr\"{o}dinger-like equation.  It is evident that there is a good agreement between them at the large separation regime.


It is also observed that the spectral structure described by Eq.(\ref{eqn5}) is composed of curves with different $\emph{n}$.  Each curve has one discontinuous point we called the branch node. The nodes exist just due to that the mathematical arctan function has the $n\pi$ ambiguity. Since ${\rm Re(A(\omega))}=0$ has a root $\omega = 1.49161$, on which the branch nodes lies.  

Spectral walls, which were investigated in \cite{AdamSpe6,AdamSWSoliton}, are defined by the point where a shape mode turns into the continuous spectrum. When a moving kink with excited vibration approaches the wall, it may pass through it or bounce back depending on the amplitude of excited shape mode. By using PBA, we successfully derive the asymptotic locations of spectral walls by taking $\omega=2$ in Equation (\ref{eqn5}), which can be expressed as 
\begin{equation}\label{eqn8}
\begin{split}
a_{sw}^{\bar{K}K}=\frac{1}{\sqrt{3}}[\arctan(\frac{{\rm Im}(A(2))}{{\rm Re}(A(2))})+(n-1)\pi],
\end{split}
\end{equation}
where \emph{n} represents the spectral wall of the \emph{n}th mode.
\begin{figure}[htp]
    \centering
    \includegraphics[width=8cm]{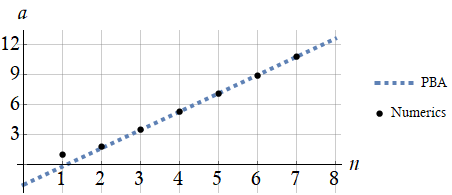}
    \caption{The locations of spectral walls versus the number of them. The dots denote numeric locations and the dashed line denotes predictions of PBA.}
    \label{SWL}
\end{figure}

In FIG.\ref{SWL}, we plot the locations of spectral walls versus the numbers of them both in PBA and numerical calculation. It indicates that the PBA has a good approximation with numerics except for the first spectral wall. Meanwhile, the difference between two adjacent locations of spectral walls  approaches to a limit. Based on Eq.(\ref{eqn8}), we could derive that this limit is ${\pi}/{\sqrt{3}}$. Thus, the PBA method enables us to predict the locations of all spectral walls in $\phi^6$ model.

By combining the dissipation modes of the single kink and their mirror images, we construct the configurations of the delocalized modes as
  \begin{equation}\label{eqn8}
\begin{split}
\delta_{s}^{\bar{K}K}(x)=&N\frac{e^{{-b^{*}(\omega)}(\left| x \right|-a)}}{(e^{-(\left| x \right|-a)}+e^{(\left| x \right|-a)})^{b(\omega)}} F(b(\omega)-\frac{3}{2},b(\omega)+\frac{5}{2},\\&k_{+}(\omega)+1,\frac{e^{-(\left| x \right|-a)}}{e^{-(\left| x \right|-a)}+e^{(\left| x \right|-a)}}),
\end{split}
\end{equation}
where (\emph{$a$}, \emph{$\omega$}) is a set of solution of Eq.(\ref{eqn5}). 

\begin{figure}[htbp]
    \centering
        \subfigure[]{
    	\includegraphics[width=8cm]{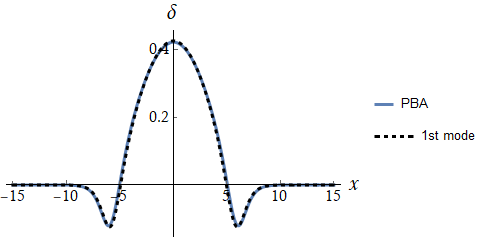}
        \label{31}
    }
    \quad    
    \subfigure[]{
        \includegraphics[width=8cm]{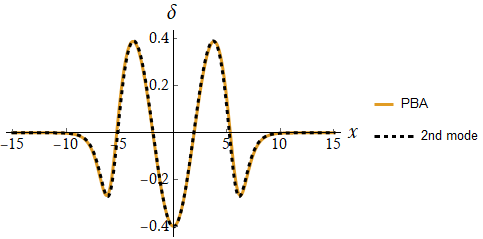}
        \label{32}
    }
        \quad    
    \subfigure[]{
        \includegraphics[width=8cm]{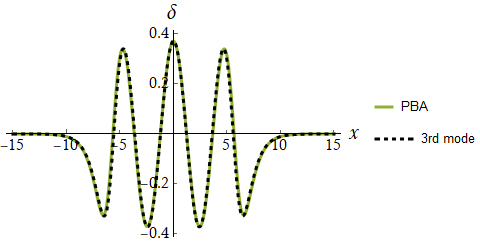}
        \label{33}
    }
        \quad    
    \subfigure[]{
        \includegraphics[width=8cm]{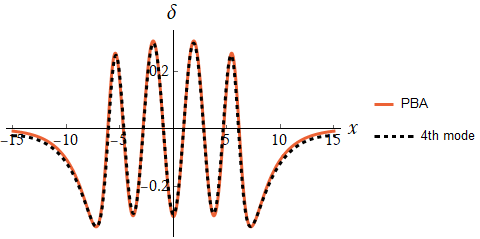}
        \label{34}
    }
    \quad
    \caption{Comparison of the configurations of four shape modes when $a=6$ in PBA(colored solid line), with numerical results of the Schr\"{o}dinger-like equation(black dashed line).}
    \label{SM}
\end{figure}

To show the applicability of PBA, we choose the half separation $a=6$ for illustration. The superposition of $\bar{K}K$ possesses four shape modes in this case. 
The first step is to put $a=6$ into Eq.(\ref{eqn5}). From solving this equation, we get four frequencies, which equals to 1.039, 1.288, 1.623 and 1.934, respectively. Then we put these frequencies into Eq.(\ref{eqn8}). After normalization, we get the corresponding configurations of shape modes. In FIG.\ref{SM}, we plot the configurations obtained by using PBA and solving the Schr\"{o}dinger-like equation numerically. The very good agreements  between them manifest the validity of the PBA method.

\section{\label{sec:level4}Kink-kink spectrum}
In the $KK$ collision ($\phi_{(-1,0)}(x+a)+\phi_{(0,1)}(x-a)$, See FIG.\ref{KK}), due to the repulsive force between kinks, no resonant bounce structure has been found \cite{Gani}. Energy transfer happens when the two kinks collides, so that it is meaningful to analyse the exited spectrum. To derive explicit forms of the excited modes, we use PBA with the same algorithm as in the $\bar{K}K$ case. 

In $KK$ collisions, the configurations of delocalized modes are odd due to the anti-symmetry of the whole configuration. Consider the  $\phi_{(0,1)}$ kink locating at $x=a$, the period boundary condition at $x=0$ can be expressed by
\begin{equation}\label{eqn9}
\begin{split}
&\frac{\partial^2\delta_{D}(x-a)}{\partial x^2}\bigg|_{x = 0} =0.
\end{split}
\end{equation}
After simplification, Eq.(\ref{eqn9}) becomes
\begin{equation}\label{eqn10}
\begin{split}
a=\frac{1}{{k_{-}(\omega)}}[\arctan(\frac{{\rm Im}(A(\omega))}{{\rm Re}(A(\omega))})+(n-\frac{1}{2})\pi],
\end{split}
\end{equation}
where \emph{n} is a positive integer.

\begin{figure}[htbp]
    \centering
        \subfigure[]{
    	\includegraphics[width=8.2cm]{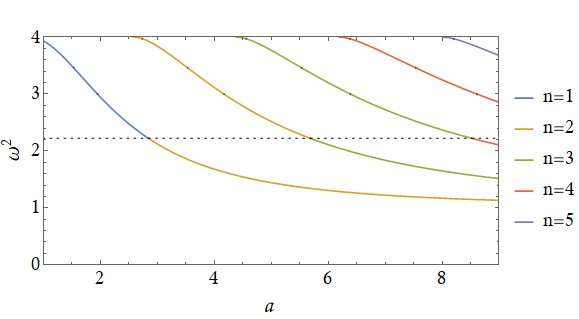}
        \label{101}
    }
    \quad    
    \subfigure[]{
        \includegraphics[width=8.7cm]{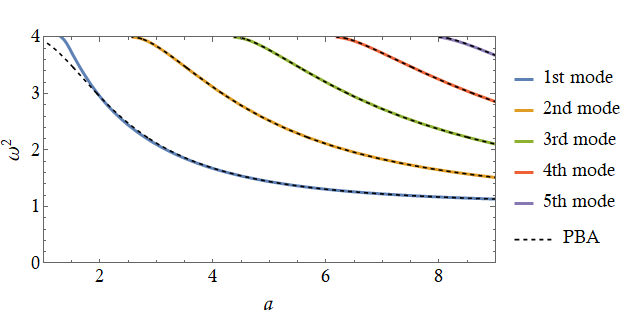}
        \label{102}
    }
        \quad    
    \caption{In the $KK$ collision: (a) Spectral structure via PBA, different colors represent curves with different \emph{n}; (b) the black dashed lines denote the merge of the colored curves in panel (a), the colored lines denote the result of the Schr\"{o}dinger-like equation.}
    \label{SM2}
\end{figure}

Compared with Eq.(\ref{eqn5}), we recognize that the only difference between them is the phase shift. The expressions of branch nodes frequencies are the same as in the $\bar{K}K$ collisions. We plot the spectral structure of the $KK$ collision in FIG.(\ref{101}) and contrast it with numerical results in FIG.(\ref{102}). We find that PBA works well in the large separation regime.

Besides, the configurations of shape modes and spectral walls of PBA fit well with numerical results by analogous tests. Here we don't go into details. We present the explicit form of locations of spectral walls below:
\begin{equation}\label{eqn11}
\begin{split}
a_{sw}^{KK}=\frac{1}{\sqrt{3}}[\arctan(\frac{{\rm Im}(A(2))}{{\rm Re}(A(2))})+(n-\frac{1}{2})\pi].
\end{split}
\end{equation}
The configuration of the delocalized mode in the $KK$ collision can be constructed analogously by combining the dissipation mode of the single kink on $x>0$ and its mirror relative to the center of the $KK$ pair. It can be expressed as
\begin{equation}\label{eqn12}
\delta_{s}^{KK}(x)=
\begin{cases}
-N\frac{e^{{b^{*}(\omega)}(x+a)}}{(e^{(x+a)}+e^{(-(x+a))})^{b(\omega)}} F(b(\omega)-\frac{3}{2},b(\omega)+\frac{5}{2},\\\quad k_{+}(\omega)+1,\frac{e^{(x+a)}}{e^{(x+a)}+e^{(-(x+a))}}),\qquad x<0,\\
N\frac{e^{{-b^{*}(\omega)}(x-a)}}{(e^{-(x-a)}+e^{(x-a)})^{b(\omega)}} F(b(\omega)-\frac{3}{2},b(\omega)+\frac{5}{2},\\\quad k_{+}(\omega)+1,\frac{e^{-(x-a)}}{e^{-(x-a)}+e^{(x-a)}}),\,\,\,\qquad x\geq0,\\
\end{cases}
\end{equation}
where (\emph{a}, \emph{$\omega$}) is a set of solution of Eq.(\ref{eqn10}).
\section{\label{sec:level5}Conclusion}
In the present work, we have found out the periodic boundary condition is the bridge that connects the localized and delocalized modes together in $\phi^6$ model for $\bar{K}K$ and $KK$ collisions. 
By using PBA, we derived the analytical forms of frequencies, spectral walls and configurations of delocalized modes. 

For potentials with the polynomial form, PBA is just valid in $\phi^6$ model. The $\phi^4$ kink does not possess the half-bound dissipation modes \cite{Manton1,Manton4}, due to the equivalent vacua with the same mass threshold. In higher order theory like $\phi^8$, the so-called long-range tails and complicate Schr\"{o}dinger potential of the single kink make PBA unable to be used\cite{Longrantail,F81,F82,F83}. 

Up to our knowledge, the effective collective coordinate approximation (CCA) in $\phi^6$ model for the $\bar{K}K$ collision constructed by \cite{Adamrev6} is not general enough due to its freezing treatment. A novel CCA needs to be proposed where the dynamics of delocalized modes are considered, and we have taken a significant step towards it. Our study will pave the way for future researchers, because analytical forms of the spectrum are more convenient than the numerical fitted results to analyse the behaviors of solitons near the spectral walls. 

\end{document}